\documentclass[twoside,11pt]{article}

\usepackage[cp1251]{inputenc}

\usepackage{graphicx}
\usepackage{amssymb}
\usepackage{amsmath}
\usepackage{amsfonts}
\usepackage{wasysym}

\begin{document}           

\title{\large SEMI-CLASSICAL UNIVERSE NEAR INITIAL SINGULARITY}
\author{V.E. Kuzmichev, V.V. Kuzmichev\\[0.5cm]
\small Bogolyubov Institute for Theoretical Physics\\
\small National Academy of Sciences of Ukraine \\ 
\small Kiev, 03680 Ukraine}

\date{}

\maketitle

\begin{abstract}
The properties of the quantum universe on extremely small spacetime scales 
are studied in the semi-classical approach to the well-defined quantum model.  
It is shown that near the initial cosmological singularity point quantum gravity effects
$\sim \hbar$ exhibit themselves in the form of additional matter source with the 
negative pressure and the equation of state as for ultrastiff matter. The analytical
solution of the equations of theory of gravity, in which matter is represented by
the radiation and additional matter source of quantum nature, is found. It is shown
that in the stage of the evolution of the universe, 
when quantum corrections $\sim \hbar$
dominate over the radiation, the geometry of the universe is described by the
metric which is conformal to a metric of a unit four-sphere in a five-dimensional 
Euclidean flat space. In the radiation dominated era the metric is found to be
conformal to a unit hyperboloid embedded in a five-dimensional Lorentz-signatured flat
space. 
The origin of the universe can be interpreted as a quantum transition of the
system from the region in a phase space with a trajectory in imaginary time 
into the region, where the equations of motion
have the solution in real time. Near the
boundary between two regions the universe undergoes
almost an exponential expansion which passes smoothly into the expansion under the
action of radiation dominating over matter. 
As a result of such a quantum transition the geometry of the
universe changes. This agrees with the hypothesis about the possible change
of geometry after the nucleation of expanding universe from `nothing'.
\end{abstract}

PACS numbers: 98.80.Qc, 04.60.-m, 04.60.Kz 

\begin{center}
      \textbf{1. Introduction}\\[0.3cm]
\end{center}

It is accepted that the present-day Universe as a whole can be considered as
a cosmological system described by the standard
model based on general relativity \cite{W,OP}. According to the Standard
Big Bang Model \cite{OP,KolT}, the early Universe was very hot and dense. In order to
describe that era one must treat the gravitational degrees of freedom and matter 
fields quantum mechanically \cite{Ish,Kie}. The fact that in the course of its 
evolution the Universe has passed through a stage with quantum degrees of freedom 
before turning into the cosmological system, whose properties are described well by 
general relativity, means that a consistent description of the Universe as a 
nonstationary cosmological system should be based on quantum general relativity in the 
form admitting the passage to general  relativity in semi-classical limit 
$\hbar \rightarrow 0$ \cite{Ish}. One of the possible versions of such a theory
with a well-defined time variable was proposed in Refs.\cite{KK1,KK2} in the case
of homogeneous, isotropic and closed universe filled with primordial matter in the
form of a uniform scalar field and relativistic matter associated with a reference
frame. As calculations have demonstrated \cite{KK1,KK2}, the equations of the
quantum model may be reduced to the form in which the matter energy density in the
universe has a component in the form of a condensate of massive quanta of a 
scalar field. Under the semi-classical description this component behaves as
an antigravitating fluid. Such a property has a quantum nature and it is connected
with the fact that the states with all possible masses of a condensate
contribute to the total wave function of the quantum universe. If one discards the 
corresponding quantum corrections, the quantum fluid degenerates into a dust, i.e.
matter component of the energy density commonly believed to make a dominant 
contribution to the mass-energy of ordinary matter in the present Universe in 
the standard cosmological model. Let us note that the presence of a condensate
in the universe, as well as the availability of a dust representing an extreme state of
a condensate, is not presupposed in the initial Lagrangian of the theory. An 
antigravitating condensate arises out of a transition from classical description 
of gravitational and matter fields to their quantum description achieved by 
canonical quantization. If one supposes that the properties of our Universe are
described in an adequate manner by such a quantum theory, an antigravitating 
condensate being found out can be associated with dark energy \cite{KK2}.
Assuming that particles of a condensate can decay to baryons, leptons (or to their
antiparticles) and particles of dark matter, one can describe the percentage of
baryons, dark matter and dark energy observed in the present Universe \cite{KK3}.

In semi-classical limit the negative pressure fluid arises as a remnant of the early
quantum era. This antigravitating component of the energy density does not
vanish in the limit $\hbar \rightarrow 0$. In addition to this component, the
stress-energy tensor contains the term vanishing after the transition to
general relativity, i.e. to large spacetime scales. However, on small spacetime
scales quantum corrections  $\sim \hbar$ turn out to be significant. As it is shown
in this paper, the effects stipulated by these corrections determine the 
equation of state of matter and geometry near the initial 
cosmological singularity point. They define a boundary condition that should be imposed
on the wave function in the origin so that a nucleation of the universe from the 
initial cosmological singularity point becomes possible.

In this paper we use the modified Planck system of units. The
$l_{P} = $ $\sqrt{2 G \hbar /(3 \pi c^{3})}$ is taken as a unit of length,
the $\rho_{P} = 3 c^{4} /(8 \pi G l_{P}^{2})$ is a unit of energy density and so on.
All relations (with the exception of Appendix A) are written for dimensionless values.

\begin{center}
      \textbf{2. Equations of motion in quantum model}\\[0.3cm]
\end{center}

Let us consider the homogeneous, isotropic and closed universe which is described
by the Robertson-Walker metric with the cosmic scale factor $a(\tau)$, where $\tau$ is
the proper time. We assume that the universe is originally filled with the uniform 
scalar field $\phi$ and a perfect fluid which defines 
so called \textit{material reference frame} \cite{KK1,BM}. 
The perfect fluid is taken in the form of relativistic matter (radiation)
with the energy density $\rho_{\gamma} = E/a^{4}$, where $E = \mbox{const}$. 
The scalar field oscillates with a small amplitude near the minimum of its 
potential energy density (potential) $V(\phi)$ at the point $\phi = \sigma$, $\left\{ 
dV(\phi)/d\phi \right\}_{\sigma} = 0$, while in general case $V(\sigma) \neq 0$. 
Then the equations of the quantum model under consideration are reduced to the 
following spectral problem \cite{KK1,KK2}
\begin{equation}\label{1}
     \left(-\partial_{a}^{2} + U_{k}(a) - E\right) f(a) = 0,
\end{equation}
where
\begin{equation}\label{2}
    U_{k}(a) =  a^{2} - 2 a M_{k} - a^{4} V(\sigma)
\end{equation}
is the effective potential, $M_{k} = m_{\sigma}(k + \frac{1}{2})$ with $m_{\sigma}^{2}
= \{d^{2}V(\phi)/d\phi ^{2}\}_{\sigma} > 0$ and $k = 0,1,2,\dots $ describes an amount
of matter (mass) in the universe in the form of a condensate of quantized scalar field
as an aggregate of massive excitation quanta of the spatially coherent oscillations of
the scalar field about the equilibrium state $\sigma$. The solutions of Eq.(\ref{1})
determine the states of the universe in an effective potential well 
(\ref{2}) at given mass of a condensate $M_{k}$. The wave function $f(a)$ and the 
eigenvalue $E$ depend on quantum number $k$, which is equal to the quantity of quanta of
a condensate, and second quantum number $n =  0,1,2,\dots $, marking the states 
(levels) in the potential well (\ref{2}). If $V(\sigma )= 0$, the spectrum of these
states is discrete \cite{KK1}. If $V(\sigma )\neq 0$, the states in the well
(\ref{2}) will be, generally speaking, quasistationary, since there exists a nonzero
probability of tunneling through the barrier into the region where its dynamics
is determined by the condition $a^{2} V(\sigma) > 1 - 2M_{k}/a$. However, if 
$V(\sigma) \ll 1$, such states can be approximately considered as stationary
within the lifetime\footnote{It can reach values comparable with the age of
our Universe \cite{KK4}.} of a system inside the barrier. The wave function may be
normalized to unity in the region limited by the barrier.

The incorporation of the reference frame through the introduction of relativistic 
matter makes it possible to define a time variable and describe the evolution of 
quantum universe by the time-dependent Schr\"{o}dinger type equation with a
time-independent Hamiltonian. The wave function of stationary states is characterized
by the parameter $E$ which has a definite value \cite{KK1,KK2}.

In the quantum model under consideration the following relation between the matrix 
elements is fulfilled \cite{KK2}
\begin{equation}\label{3}
    \langle f | - \frac{i}{N}\, \frac{d}{d\eta}\,\partial_{a}|f \rangle = 
     \langle f \left|\left(a - 2a^{3} V(\sigma) - 4M_{k}\right)\right| f \rangle 
\qquad \mbox{for} \qquad k \gg 1,
\end{equation}
where $\eta$ is the time variable which is connected with the proper time $\tau$ by 
the differential equation $d\tau = a N d\eta$, $N = (g^{00})^{-1/2}$ 
is the lapse function 
that specifies the time reference scale \footnote{Even for the Planck mass 
$M_{k} = m_{P}  = 1$ the number $k \sim 10^{18}$, if 
$m_{\sigma} \sim 10^{-18}(\sim 10 \ \mbox{GeV})$ \cite{KK1}.}.

Choosing the function $f(a)$ in the form
\begin{equation}\label{4}
    f(a) = A(a)\, \mbox{e}^{i S(a)},
\end{equation}
where $A$ and $S$ are real functions of $a$,
and assuming that the matter-energy in semi-classical universe can be written 
in the form of perfect fluid source \cite{OP} 
we obtain from (\ref{1}) and (\ref{3}) the equations for $A$ and $S$
\begin{equation}\label{5}
   \frac{1}{a^{4}} \left(\partial_{a} S \right)^{2} - \rho  + \frac{1}{a^{2}} - 
\frac{1}{a^{4}}\,\frac{\partial_{a}^{2} A}{A}-
\frac{i}{a^{4}}\,\frac{\partial_{a} \left(A^{2} \partial_{a} S \right)}{ A^{2}} = 0,
\end{equation}
\begin{equation}\label{6}
   \frac{1}{a^{2}}\,\frac{d}{d \tau}\,\left(\partial_{a} S \right) + 
\frac{1}{2}\,\left(\rho - 3 p \right) - \frac{1}{a^{2}} -
\frac{i}{a^{2}}\,\frac{d}{d \tau}\,\left(\frac{\partial_{a} A}{A} \right) = 0,
\end{equation}
where
\begin{equation}\label{7}
    \rho = \rho_{k} +\rho_{\sigma} + \rho_{\gamma}, \qquad 
     p = p_{k} + p_{\sigma} + p_{\gamma}
\end{equation}
are the energy density and the isotropic pressure as the sum of the components
\begin{equation}\label{8}
    \rho_{k} = \frac{2 M_{k}}{a^{3}}, \qquad  \rho_{\sigma} \equiv V(\sigma) \equiv 
\frac{\Lambda}{3}, \qquad \rho_{\gamma} = \frac{E}{a^{4}},
\end{equation}
$\Lambda$ is the cosmological constant. The equations of state are
\begin{equation}\label{9}
   p_{k} = - \rho_{k}, \qquad  p_{\sigma} = - \rho_{\sigma},\qquad p_{\gamma} = 
\frac{1}{3}\,\rho_{\gamma}. 
\end{equation}
The equations of state for the vacuum component $\rho_{\sigma} = \mbox{const.}$ and 
relativistic matter $\rho_{\gamma}$ are dictated by the formulation of the problem. 
The vacuum-type equation of state of a condensate with the density $\rho_{k}$, which
does not remain constant throughout the evolution of the universe, but decreases 
according to a power law with the increase of $a$, follows from the 
condition of consistency of Eqs. (\ref{5}) and (\ref{6}).

From Eqs. (\ref{5}) - (\ref{9}) we can conclude that a condensate behaves as an 
anti-gravitating medium. Its anti-gravitating effect has a purely quantum nature. 
Its appearance is determined by the fact that the total wave function of the universe
$\psi = \sum_{k} f_{k} u_{k}$ is a superposition of quantum states
with all possible values of the quantum number $k$, $u_{k}$ is the wave function, 
which describes the quantum states of a scalar field. In the quantum model with the
potential $V(\phi )$, which has a harmonic oscillator form 
near the point $\phi = \sigma$, the contribution into the right-hand side of 
Eq. (\ref{3}) is made by the states with $k$ and $k\pm 2$ only. If one discards
the contributions from the states with $k\pm 2$, a condensate turns into
an aggregate of separate macroscopic bodies with zero
pressure (dust) \cite{KK2}. The existence of this limit argues in favour
of reliability of this quantum model.

In the classical limit $(\hbar = 0)$, when all terms with the amplitude $A$
are discarded (see Appendix A), Eqs. (\ref{5}) and (\ref{6}) reduce to
the Einstein-Friedmann equations which predict an accelerating expansion
of the universe in the era with $\rho_{k} > \frac{2}{3}\,\rho_{\gamma}$, even if 
$\Lambda = 0$. Since $\rho_{\gamma} \sim a^{-4}$ decreases with $a$ more rapidly
then $\rho_{k} \sim a^{-3}$ (or even $\sim a^{-2}$ \cite{KK2}), the era
of accelerating expansion should begin with increasing $a$, even if 
the state with $\rho_{k} < \frac{2}{3} \,\rho_{\gamma}$ and $\Lambda \sim 0$ 
existed in the past, when the expansion was decelerating.

\begin{center}
      \textbf{3. Semi-classical universe}\\[0.3cm]
\end{center}

Separating in Eq. (\ref{5}) the real and imaginary parts and setting each one of them 
equal to zero separately, from the equation for the imaginary part 
$\partial_{a} \left(A^{2} \partial_{a} S \right) = 0$, we find the
amplitude $A$,
\begin{equation}\label{10}
    A = \frac{\mbox{const.}}{\sqrt{\partial_{a}S}}\,.
\end{equation}
Substituting this solution into the real part of Eq. (\ref{5}) and into Eq. (\ref{6}), 
we arrive at the equations for the phase $S$,
\begin{equation}\label{11}
   \frac{1}{a^{4}} \left(\partial_{a} S \right)^{2} - \rho  + \frac{1}{a^{2}} = 
\frac{1}{a^{4}}\,\left\{\frac{3}{4}\, \left(\frac{\partial_{a}^{2} 
S}{\partial_{a}S}\right)^{2} - \frac{1}{2}\,\frac{ \partial_{a}^{3} 
S}{\partial_{a}S}\right\},
\end{equation}
\begin{equation}\label{12}
   \frac{1}{a^{2}}\,\frac{d}{d \tau}\,\left(\partial_{a} S \right) + 
\frac{1}{2}\,\left(\rho - 3 p \right) - \frac{1}{a^{2}} = -\, 
\frac{i}{2 a^{2}}\,\frac{d}{d \tau}\,\left(\frac{\partial_{a}^{2} S}{\partial_{a}S} 
\right).
\end{equation}
These equations are exact. If the solution of these nonlinear equations is found, 
it would be possible, theoretically, to
restore the wave function (\ref{4}). Rewriting Eqs. (\ref{11}) and (\ref{12}) in 
ordinary units (see Appendix A), we find that the right-hand sides of Eqs. (\ref{11})
and (\ref{12}) are proportional to $\hbar^{2}$ and $\hbar$ respectively. 
It means that the right-hand sides of these equations can be considered as small
quantum corrections to the equations of general relativity.

Quantum corrections are essential in the region of extremely small values of the scale
factor. Substituting the standard model solution of the Einstein-Friedmann equations
(see Appendix B)
\begin{equation}\label{13}
    a(\tau) = \beta\, \tau^{\alpha},
\end{equation}
$\alpha > 0$ and $\beta> 0$ are constants, 
into the expression for the generalized momentum
\begin{equation}\label{14}
    \partial_{a} S = - a\, \frac{da}{d\tau} \equiv - a\dot{a},
\end{equation}
we find that this solution satisfies Eqs. (\ref{11}) and (\ref{12}) in the region,
where quantum corrections play a main role \footnote{See Eq. (\ref{30}) below.}.
Using Eq. (\ref{13}) in order to calculate the right-hand sides of 
Eqs. (\ref{11}) and (\ref{12}) we obtain the following equations
\begin{equation}\label{15}
   \frac{1}{a^{4}} \left(\partial_{a} S \right)^{2} = 
\rho  + \frac{d_{\alpha}}{a^{6}} - \frac{1}{a^{2}}, 
\end{equation}
\begin{equation}\label{16}
   \frac{1}{a^{2}}\,\frac{d}{d \tau}\,\left(\partial_{a} S \right) + 
\frac{1}{2}\,\left(\rho - 3 p \right) - \frac{1}{a^{2}} = -\, i\,
\frac{b_{\alpha}}{2 a^{5}}\,\partial_{a} S,
\end{equation}
where we denote
\begin{equation}\label{17}
    d_{\alpha} \equiv \frac{(2 \alpha - 1)(4\alpha - 1)}{4 \alpha^{2}}, \qquad
    b_{\alpha} \equiv \frac{2 \alpha - 1}{\alpha}.
\end{equation}
These equations take into account the presence of matter with the energy density
$\rho$ and describe the evolution of the universe in the semi-classical 
approximation. 

Eq. (\ref{15}) can be considered as the first of the Einstein-Friedmann equations
for the matter which is characterized by the generalized energy density 
$\rho + d_{\alpha} a^{-6}$. It means that the quantum effects under consideration 
cause the appearance of an additional source of gravitational field which decreases
with $a$ as $a^{-6}$. If we regard Eq. (\ref{16}) as the second of the 
Einstein-Friedmann equations with the generalized energy density and pressure,
then comparing the quantum corrections with the energy density of the standard model
(see Appendix B), we find that it should be set $\alpha = \frac{1}{3}$, and
$d_{1/3} = - \frac{1}{4}$, $b_{1/3} = - 1$.

Using the representation for the classical momentum (\ref{14}) Eqs. (\ref{15}) 
and (\ref{16}) can be reduced to the standard form
\begin{equation}\label{18}
    \left(\frac{\dot{a}}{a}\right)^{2} = \rho + \rho_{u} - 
    \frac{1}{a^{2}},
\end{equation}
\begin{equation}\label{19}
    \frac{\ddot{a}}{a} = 
     -\,\frac{1}{2}\,\left[\rho + \rho_{u} + 3( p + p_{u})\right],
\end{equation}
where 
\begin{equation}\label{20}
    \rho_{u} = - \frac{1}{4a^{6}}
\end{equation}
is the quantum correction ($\sim \hbar^{2}$) 
\footnote{Let us note that the presence of a minus sign in Eq. (\ref{20}) is not
extraordinary. So, according to quantum field theory, for instance, vacuum fluctuations
make a negative contribution to the field energy per unit area (the Casimir effect).}, 
which may be identified with the ultrastiff matter with the equation of state
 \footnote{Specifying the form of the equation of state for the density $d_{\alpha} 
a^{-6}$ as $p_{u} = w \rho_{u}$, where $w$ is constant which has to be found, and 
performing the corresponding calculations we obtain the equations
from which it follows that  $\alpha = \frac{1}{3}$ and $w=1$.}
\begin{equation}\label{21}
    p_{u} = \rho_{u},
\end{equation}
where $p_{u}$ is the pressure. This `matter' has quantum origin. It is
formed due to nonzero derivatives of the amplitude $A$ of the wave function
(\ref{4}) and is related nonlinearly with matter components contributing into
the energy density $\rho$ (\ref{7}).

Let us note that when deriving the Eq. (\ref{19}) from Eq. (\ref{16}) we have used
Eq. (\ref{15}), while in the right-hand side of Eq. (\ref{16}) only the main term in the
range of action of the quantum correction, $2 E a^{2} < 1$, is taken into account.
In the region $2 E a^{2} > 1$ the density $\rho_{u}$, and pressure $p_{u}$ may be 
neglected. In this case equations correspond to the limit 
$\hbar \rightarrow 0$ \cite{KK2}. They coincide formally with the equations
of standard cosmology, but, unlike them, besides the familiar contributions
into the energy density  $\rho$ (\ref{7}) from the vacuum term $\rho_{\sigma}$ and 
radiation $\rho_{\gamma}$, it contains a nontrivial contribution from
antigravitating quantum fluid with the energy density $\rho_{k}$ which
does not vanish at $\hbar \rightarrow 0$.

Let us estimate the ratio of energy density $\rho$ to $\rho_{u}$. Passing
to the ordinary units (see Appendix A), we have
\begin{equation}\label{22}
    \mathcal {R} \equiv \left[ \frac{8\pi G}{3c^{4}}\,\rho \right] : 
     \left[\left(\frac{2 G \hbar}{3\pi c^{3}}\right)^{2} \frac{1}{4 a^{6}}\right].
\end{equation}
Substituting here the values of the fundamental constants we obtain
\begin{equation}\label{23}
    \mathcal {R} \sim 10^{81}\, \rho \, a^{6},
\end{equation}
where $\rho$ is measured in GeV/cm$^{3}$ and $a$ in cm. 
For our Universe today $\rho \sim 10^{-5}$ GeV/cm$^{3}$, $a \sim 10^{28}$ cm and
\begin{equation}\label{24}
    \mathcal {R}_{today} \sim 10^{244},
\end{equation}
i.e. the quantum correction may be neglected to an accuracy of $\sim O(10^{-244})$. 
In the Planck
era $\rho \sim 10^{117}$ GeV/cm$^{3}$, $a \sim 10^{-33}$ cm and the relation
\begin{equation}\label{25}
    \mathcal {R}_{Planck} \sim 1
\end{equation}
shows that the densities $\rho$ and $\rho_{u}$ are of the same order of magnitude.

In Fig.~1 the energy density $\rho + \rho_{u}$ as a function of $a$ for typical values
of the parameters of the early universe is shown. It is evident that the range of 
action of the quantum correction is limited by the condition 
$a \lesssim 0.5$. A small variation of the parameters does not affect the result.

\begin{figure}[ht]
\begin{center}
\includegraphics*{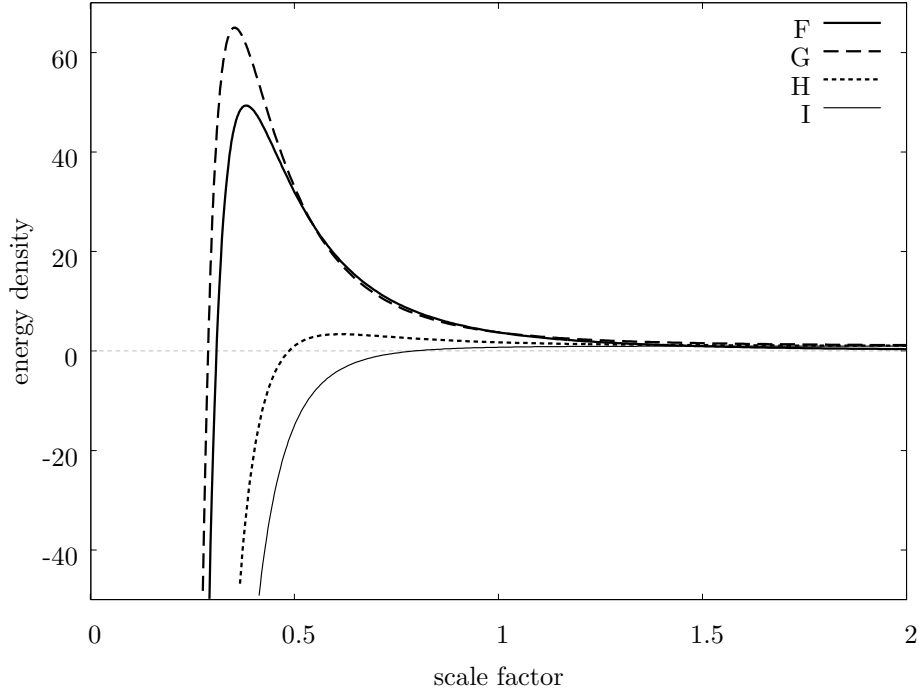}
\end{center}
\caption{The energy density $\rho + \rho_{u}$ vs. the scale factor $a$. 
The following parameters are used: 
$F = \{M_{k} = 1,\,\rho_{\sigma} = 0,\,E = 2 \}$, 
$G = \{M_{k} = 0,\,\rho_{\sigma} = 1,\,E = 1 \}$,
$H = \{M_{k} = 0,\,\rho_{\sigma} = 1,\,E = 3 \}$,
$I = \{M_{k} = 0,\,\rho_{\sigma} = 1,\,E = 0 \}$.
The curves $G$, $H$, and $I$ correspond to the potential (\ref{2}) with
$M_{k} = 0$. The $I$ describes the standard model with nonzero cosmological constant in
the absence of matter.} \label{fig:1}
\end{figure}

\begin{center}
      \textbf{4. Quantum effects on sub-Planck scales}\\[0.3cm]
\end{center}

Let us consider the solutions of Eqs. (\ref{18}) and (\ref{19}) in the region
$2 E a^{2} < 1$. Here the contributions from the condensate, cosmological constant 
and curvature may be neglected. As a result the equations of the model take the form
\begin{equation}\label{26}
    \frac{1}{2}\,\dot{a}^{2} + U(a) = 0,\qquad 
          \ddot{a} = - \frac{dU}{da},
\end{equation}
where
\begin{equation}\label{27}
    U(a) \equiv \frac{1}{2}\, \left[ \frac{1}{4 a^{4}} - \frac{E}{a^{2}} \right].
\end{equation}
These equations are similar to ones of Newtonian mechanics. Using this analogy
they can be considered as equations which describe the motion of a `particle'
with a unit mass and zero total energy under the action of the force
$-\frac{dU}{da}$, $U(a)$ is the potential energy, and 
$a(\tau)$ is a generalized variable.
A point $a_{c} = \frac{1}{2 \sqrt{E}}$, where $U(a_{c}) = 0$, divides the region of 
motion of a `particle' into the subregion $4 E a^{2} < 1$, where the classical motion of 
a `particle' is forbidden, and the subregion  $4 E a^{2} > 1$, where the classical
trajectory of a `particle' moving in real time $\tau$ exists. The quantum corrections
are essential in the region  $2 E a^{2} < 1$. The boundary point 
$a_{b} = \frac{1}{\sqrt{2 E}}$ corresponds to the minimum of the potential energy
(\ref{27}). The solution can be extended to the full range of values of the variable
$a > a_{b}$.

In the subregion $4 E a^{2} < 1$ there exists the classical trajectory of a `particle' 
moving in imaginary time $t = - i \tau + \mbox{const}.$ in the potential $- U(a)$.
Denoting the corresponding solution as $\mathrm{a}$ we find
\begin{equation}\label{28}
   \mathrm{a} = \frac{1}{2 \sqrt{E}}\, \sin z,
\end{equation}
\begin{equation}\label{29}
    t = \frac{1}{16 E^{3/2}}\, [2 z - \sin 2z].
\end{equation}
At small $z$, i.e. in the region $\mathrm{a} \sim 0$, we have
\begin{equation}\label{30}
    \mathrm{a} = \left(\frac{3}{2}\,t \right)^{1/3}.
\end{equation}
According to the standard model solution (see Appendix B) the `matter' near the
point $\mathrm{a} = 0$ is described by the equation of state of the ultrastiff matter
(\ref{21}).

In the subregion $4 E a^{2} > 1$ the solution of the equations (\ref{26}) 
can be written as
\begin{equation}\label{31}
    a = \frac{1}{2 \sqrt{E}}\,\cosh \zeta,
\end{equation}
\begin{equation}\label{32}
    \tau = \frac{1}{16 E^{3/2}}\, [2 \zeta + \sinh 2\zeta].
\end{equation}
At $\zeta \ll 1$ it follows from here that $\zeta \approx 4 E^{3/2} \tau$, while
the scale factor at $\tau \ll (4 E^{3/2})^{-1}$ increases almost exponentially
\begin{equation}\label{33}
    a = \frac{1}{2 \sqrt{E}}\,\left[1 + (2 E)^{3} \tau^{2} + \ldots \right]
       \approx \frac{1}{2 \sqrt{E}}\, \exp \left\{(2 E)^{3} \tau^{2} \right\}. 
\end{equation}

The almost exponential expansion of the early universe in that era is
stipulated by the action of quantum effects which, according to Eqs. (\ref{20})
and (\ref{21}), cause the negative pressure, $p_{u} < 0$, i.e. produce an 
anti-gravitating effect on the cosmological system under consideration.

At $\zeta \gg 1$ the solution (\ref{31}), (\ref{32}) takes the form
\begin{equation}\label{34}
    a = \left(2 \sqrt{E}\, \tau \right)^{1/2}.
\end{equation}
It describes the radiation dominated era and corresponds to the time $\tau$ which
satisfies the condition $\frac{1}{2}\,\ln (32 E^{3/2} \tau) \gg 1$.

The solutions (\ref{28}), (\ref{29}) and (\ref{31}), (\ref{32}) are related between 
themselves through an analytic continuation into the region of complex values
of the time variable,
\begin{equation}\label{35}
    t = - i \tau + \frac{\pi}{16 E^{3/2}}, \qquad z = \frac{\pi}{2} - i \zeta.
\end{equation}
The scale factors (\ref{28}) and (\ref{31}) are connected through the condition
\begin{equation}\label{36}
    a(\tau) = \mathrm{a} \left(\frac{\pi}{16 E^{3/2}} - i \tau \right),
\end{equation}
which describes an analytic continuation of the time variable $\tau$ into the region
of complex values of Euclidean time $t$.

The model determined by the equations (\ref{26}) allows us to describe the origin 
(nucleation) of the universe as the transition from the state in the subregion
$4 E a^{2} < 1$ to the state in the subregion $4 E a^{2} > 1$. The corresponding 
transition amplitude can be written as follows \cite{Col}
\begin{equation}\label{37}
    T \sim e^{-S_{t}},
\end{equation}
where $S_{t}$ is the action on a trajectory in imaginary time $t$,
\begin{equation}\label{38}
    S_{t} = 2 \int_{-\infty}^{\infty}\!\!dt U(\mathrm{a}).
\end{equation}
Let us proceed to the integration with respect to the time variable $z$. According
to (\ref{28}) the scale factor $\mathrm{a}$ is a periodical function of $z$. Therefore
we shall restrict ourselves to the interval of integration
$z = [-\frac{\pi}{2}, \frac{\pi}{2}]$. On the boundaries of this interval the 
$|\mathrm{a}|$ takes the maximum possible value $a_{c} = \frac{1}{2 \sqrt{E}}$,
while the interval itself contains the point $\mathrm{a} = 0$. Then
\begin{equation}\label{39}
    S_{t} = 2 \int_{-\frac{\pi}{2}}^{\frac{\pi}{2}}\!\!
       dz \frac{dt}{dz} U(\mathrm{a}(z)).
\end{equation}
Using the explicit form of the solution (\ref{28}), (\ref{29}), we find
\begin{equation}\label{40}
    S_{t} = - \sqrt{E} \pi,
\end{equation}
and the amplitude (\ref{37}) becomes
\begin{equation}\label{41}
    T \sim e^{\sqrt{E} \pi},
\end{equation}
i.e. a `particle' which is the equivalent of the universe leaves the subregion forbidden
for classical motion with an exponential probability density. It is pushed out of 
forbidden subregion into the subregion of very small values of $a$ in real time $\tau$
by the anti-gravitating forces stipulated by the negative pressure which cause 
quantum processes at $a \sim 0$ (see Eqs. (\ref{20}) and (\ref{21})). This phenomenon
can be interpreted as the origin of the universe from the region $a \sim 0$. It is
possible only if the probability density that the universe is in the state with
$a \sim 0$ is nonzero. It means that the wave function of the universe at the point
$a = 0$ must be nonvanishing.

\begin{center}
      \textbf{5. Wave function near initial singularity}\\[0.3cm]
\end{center}

Let us determine the behaviour of the wave function $f(a)$ in the region $a \sim 0$
and calculate the nucleation rate of the universe from the point $a = 0$.
The result will be compared with the transition amplitude (\ref{41}).
 
From Eqs. (\ref{4}) and (\ref{10}) it follows that
\begin{equation}\label{42}
    f(a) = \frac{f_{0}}{\sqrt{\partial_{a}S_{E}}}\,\,\mbox{e}^{-S_{E}},
\end{equation}
where $f_{0} = \mbox{const.}$, $S_{E} = - i S$ is the Euclidean action.
Using Eqs. (\ref{14}) and (\ref{30}) we find
\begin{equation}\label{43}
    \partial_{a}S_{E} = \frac{1}{2a}.
\end{equation}
The integration of this equation gives
\begin{equation}\label{44}
    S_{E} =  \frac{1}{2}\,\ln \left (\frac{a}{a_{0}}\right ),
\end{equation}
where the integration constant is taken in the form $\ln a_{0}^{-1/2}$, 
for convenience. Then the wave function
\begin{equation}\label{45}
    f(a) = \sqrt{2\,a_{0}}\,f_{0}
\end{equation}
does not depend on $a$ in the region $a \sim 0$, and the probability density that the
universe may be found at the point $a = 0$ is nonzero,
\begin{equation}\label{46}
    |f(0)|^{2} = 2\,a_{0}\,|f_{0}|^{2}.
\end{equation}

The nucleation rate of the universe from the initial cosmological singularity point
$a = 0$ can be written as follows \cite{KK1}
\begin{equation}\label{47}
    \Gamma = \overline{v\,\sigma_{r}}\, |f(0)|^{2},
\end{equation}
where the multiplier $\overline{v\,\sigma_{r}}$ has to be calculated now with 
respect to imaginary time $t$,
\begin{equation}\label{48}
    \overline{v\,\sigma_{r}} = \lim_{a \rightarrow 0} \left(\frac{da}{dt}\,
     \pi a^{2}\right) = \frac{\pi}{2}.
\end{equation} 
Taking into account Eq. (\ref{46}) we obtain
\begin{equation}\label{49}
    \Gamma = \pi\,a_{0}\, |f_{0}|^{2}.
\end{equation}
The constant $a_{0}\,|f_{0}|^{2}$ can be found 
as a result of exact integration of Eq. (\ref{1}) with the
boundary condition $f(0) = \mbox{const.}$, in accordance with Eq. (\ref{45}). 
For the case  
$\rho_{\sigma} = 0$  we obtain that a rate of nucleation of the universe in
the $n$-state follows the law \cite{KK1}
\begin{equation}\label{50}
    \Gamma \simeq \frac{\sqrt{\pi}}{2}\, 2^{n}\, \mathcal{P}(n),
\end{equation}
where $\mathcal{P}(n) = (\langle n \rangle^{n}/n!)\, \exp(-\langle n \rangle^{n})$ is
the Poisson distribution with the mean value $\langle n \rangle = M_{k}^{2}$.
Summing over all values of the quantum number $n$ the total nucleation rate appears to be exponentially high
\begin{equation}\label{51}
    \Gamma _{tot} \simeq \frac{\sqrt{\pi}}{2}\,\mbox{e}^{M_{k}^{2}}.
\end{equation}
Comparing the nucleation rate (\ref{51}) with the transition amplitude (\ref{41})
we see that both these quantities predict an exponential origin (nucleation) of
the universe from the region $a \sim 0$ forbidden for classical motion.

\begin{center}
      \textbf{6. Geometry }\\[0.3cm]
\end{center}

Let us consider how the geometry of the universe changes as a result of its 
transition from the region $2 E a^{2} < \frac{1}{2}$ 
into $\frac{1}{2} < 2 E a^{2} < 1$.
In the model under consideration the metric has the form
\begin{equation}\label{52}
    ds^{2} = d\tau^{2} - a^{2} d\Omega_{3}^{2},
\end{equation}
where $d\Omega_{3}^{2}$ is a line element on a unit three-sphere. 
According to the solutions (\ref{28}), (\ref{29}) and (\ref{31}), (\ref{32}) the
metric (\ref{52}) takes the form
\begin{equation}\label{53}
    ds_{E}^{2} = - \frac{1}{4 E}\,\sin^{2} z \left\{\frac{\sin^{2} z}{4 E^{2}}\,
        dz^{2} + d\Omega_{3}^{2} \right\} \quad \mbox{at} \quad 2 E a^{2} < \frac{1}{2}
\end{equation}
and
\begin{equation}\label{54}
    ds_{L}^{2} = \frac{1}{4 E}\,\cosh^{2} \zeta \left\{\frac{\cosh^{2} \zeta}{4 E^{2}}\,
        d\zeta^{2} - d\Omega_{3}^{2} \right\} 
        \quad \mbox{at} \quad \frac{1}{2} < 2 E a^{2} < 1,
\end{equation}
where the interval with the Euclidean signature is denoted by the index $E$, and the one
with the Lorentzian signature is marked by $L$. Introducing the new time variables $\xi$
and $\varsigma$ according to
\begin{equation}\label{55}
    d\xi = \frac{1}{2 E} \sin z dz, 
        \qquad d\varsigma = \frac{1}{2 E} \cosh \zeta d\zeta,
\end{equation}
the metrics (\ref{53}) and (\ref{54}) can be reduced to the conformally flat form
\begin{equation}\label{56}
    ds_{E}^{2} = - \frac{1}{4 E} \left[1 - (2 E \xi)^{2} \right]
        \left\{d\xi^{2} + d\Omega_{3}^{2} \right\},
\end{equation}
\begin{equation}\label{57}
    ds_{L}^{2} = \frac{1}{4 E} \left[1 + (2 E \varsigma)^{2} \right]
        \left\{d\varsigma^{2} - d\Omega_{3}^{2} \right\}.
\end{equation}
Both metrics are related between themselves through the analytic continuation
into the region of complex values of the time variable $\varsigma = i \xi$.

The metric (\ref{56}) is conformal to a metric of a unit four-sphere in a 
five-dimensional Euclidean flat space. With increasing $a$, the universe transits
from the region $4 E a^{2} < 1$ into the region $4 E a^{2} > 1$,
where the geometry is conformal to a unit hyperboloid embedded in a 
five-dimensional Lorentz-signatured flat space.
Such a picture of change 
in spacetime geometry during the transition of the universe from the
region near initial singularity into the region of real physical scales agrees with the
hypothesis \cite{HH,HP}, widely discussed in the literature 
for the de Sitter space, 
about possible change in four-space 
geometry after the spontaneous nucleation of the expanding universe from the initial
singularity point.
In this paper this phenomenon is demonstrated in the case of the early universe
filled with the radiation and ultrastiff matter which effectively takes into account
quantum effects on extremely small spacetime scales. 

\begin{center}
      \textbf{7. Conclusions }\\[0.3cm]
\end{center}

In this paper we study the properties of the quantum universe on extremely small 
spacetime scales in the semi-classical approach to the well-defined quantum model. 
We show that near the initial cosmological singularity point quantum gravity effects
$\sim \hbar$ exhibit themselves in the form of additional matter source with the 
negative pressure and the equation of state as for ultrastiff matter. The analytical
solution of the equations of theory of gravity, in which matter is represented by
the radiation and additional matter source of quantum nature, is found. It is shown
that in the stage of the evolution of the universe, 
when quantum corrections $\sim \hbar$
dominate over the radiation, the geometry of the universe is described by the
metric which is conformal to a metric of a unit four-sphere in a five-dimensional 
Euclidean flat space. In the radiation dominated era the metric is found to be
conformal to a unit hyperboloid embedded in a five-dimensional Lorentz-signatured flat
space. One solution can be continued analytically into another.

The wave function of the universe in the initial cosmological singularity point is 
nonzero and the nucleation of the universe from this point becomes possible.
The origin of the universe can be interpreted as a quantum transition of the
system from the region in a phase space forbidden for classical motion, but where
a trajectory in imaginary time exists, into the region, where the equations of motion
have the solution which describes the evolution of the universe in real time. Near the
boundary between two regions, from the side of real time, the universe undergoes
almost an exponential expansion which passes smoothly into the expansion under the
action of radiation dominating over matter which is described by the standard
cosmological model. As a result of such a quantum transition the geometry of the
universe changes. This agrees with the hypothesis about the possible change
of geometry after the nucleation of expanding universe from `nothing'.

\begin{center}
      \textbf{Appendix A }\\[0.3cm]
\end{center}

In the ordinary units the wave function (\ref{4}) has the form
$$
  f(a) = A(a)\, \exp \left\{\frac{i}{\hbar}\,\frac{3\pi c^{3}}{2G}\,S(a)\right\},
$$
where $S$ is measured in cm$^{2}$. Eqs. (\ref{11}) and (\ref{12}) can be 
written as follows
$$
   \frac{1}{a^{4}} \left(\partial_{a} S \right)^{2} - \frac{8\pi G}{3 c^{4}}\,\rho  + 
\frac{1}{a^{2}} = \left(\frac{2G\hbar}{3 \pi c^{3}}\right)^{2}
\frac{1}{a^{4}}\,\left\{\frac{3}{4}\, \left(\frac{\partial_{a}^{2} 
S}{\partial_{a}S}\right)^{2} - \frac{1}{2}\,\frac{ \partial_{a}^{3} 
S}{\partial_{a}S}\right\},
$$
$$
   \frac{1}{a^{2}}\,\frac{d}{d \tau}\,\left(\partial_{a} S \right) + 
\frac{4 \pi G}{3 c^{4}}\,
\left(\rho - 3 p \right) - \frac{1}{a^{2}} = - \frac{2 G \hbar}{3 \pi c^{3}}
\,\frac{i}{2 a^{2}}\,\frac{d}{d \tau}\,\left(\frac{\partial_{a}^{2} S}{\partial_{a}S} 
\right).
$$
Here $a$ is measured in cm, $\rho$ and $p$ in GeV/cm$^{3}$, and $c^{4}/G$ in GeV/cm.
In the approximation $\hbar = 0$ these equations strictly pass into the equations of general relativity for generalized momentum $\partial_{a} S = - a\, da/d\tau$.         

\begin{center}
      \textbf{Appendix B }\\[0.3cm]
\end{center}

As it is well known \cite{OP,KolT} the standard model solution of the 
Einstein-Friedmann 
equations for a single component domination in the energy density $\rho$ have the form
$$
   \rho \sim a^{-3(1+w)}
$$
for the equation of state $w = p /\rho = \mbox{const}$. If $w\neq -1$
and the curvature term can be neglected
the scale factor $a$ is
$$
   a\sim \tau^{2/[3(1+w)]}.
$$
According to accepted notations \cite{KolT,DF} 
the value $w = -\frac{2}{3}$ describes the 
domain walls ($\rho \sim a^{-1}, \, a \sim \tau^{2}$); the value $w= -\frac{1}{3}$
corresponds to the strings ($\rho \sim a^{-2}, \, a \sim \tau $); $w= 0$ is a dust
($\rho \sim a^{-3}, \, a \sim \tau^{2/3}$), $w= \frac{1}{3}$ is a radiation 
($\rho \sim a^{-4}, \, a \sim \tau^{1/2}$); $w= \frac{2}{3}$ is a perfect gas
($\rho \sim a^{-5}, \, a \sim \tau^{2/5}$), and $w= 1$ is an ultrastiff matter
($\rho \sim a^{-6}, \, a \sim \tau^{1/3}$). The special case with 
$w = - 1$ and $\rho = \mbox{const.}$ describes the de Sitter vacuum.


\begin{thebibliography}{99}
\itemsep -6pt plus 1pt minus 1pt
\bibitem{W}R.M. Wald, \textit{General Relativity}, University of Chicago Press, 
Chicago and London 1984.

\bibitem{OP}K.A. Olive, T.A. Peacock, \textit{Phys. Lett.} \textbf{B667}, 217 (2008).

\bibitem{KolT}E.W. Kolb, M.S. Turner, \textit{The Early Universe}, 
Addison-Wesley Publishing Company, Redwood City 1990.

\bibitem{Ish}C.J. Isham, gr-qc/9510063.

\bibitem{Kie}K. Kiefer, \textit{Lecture Notes in Physics 541: Towards
Quantum Gravity}, ed. J. Kowalski-Glikman, Springer-Verlag, Heidelberg 
2000 [gr-qc/9906100].

\bibitem{KK1}V.E. Kuzmichev, V.V. Kuzmichev, \textit{Acta Phys. Pol. B} 
\textbf{39}, 979 (2008) [gr-qc/0712.0464].

\bibitem{KK2}V.E. Kuzmichev, V.V. Kuzmichev, \textit{Acta Phys. Pol. B} 
\textbf{39}, 2003 (2008) [gr-qc/0712.0465].

\bibitem{KK3}V.E. Kuzmichev, V.V. Kuzmichev, in: \textit{Trends in Dark Matter 
Research}, ed. J.V. Blain, Nova Science, Hauppage 2005 [astro-ph/0405455].

\bibitem{BM}J.D. Brown, D. Marolf, \textit{Phys. Rev.} \textbf{D53}, 
1835 (1996) [gr-qc/9509026].

\bibitem{KK4}V.E. Kuzmichev, V.V. Kuzmichev, \textit{Eur. Phys. J.} \textbf{C23},
337 (2002) [astro-ph/0111438];
\textit{Ukr. J. Phys.} \textbf{51}, 433 (2006).

\bibitem{DF}I. Dymnikova, M.Fil'chenkov, \textit{Phys. Lett.} \textbf{B545}, 214 
(2002).

\bibitem{Col}S. Coleman, \textit{Phys. Rev.} \textbf{D15}, 2929 
(1977).

\bibitem{HH}J.B.Hartle, S.W. Hawking, \textit{Phys. Rev.} \textbf{D28}, 2960 (1983).

\bibitem{HP}S.Hawking, R Penrose, \textit{The Nature of Space and Time}, Princeton 
University Press, New Jersey 2000.






\end{thebibliography}
\end{document}